\documentclass{emulateapj}
\usepackage{amsmath}
\usepackage{epsfig,natbib}
\usepackage{amsmath,amsfonts,amssymb}
\usepackage{mathptmx}
\usepackage{color}
\usepackage{graphicx}
\slugcomment{}

\shorttitle{An Ultra Low Mass and Small Radius Compact Object in \object{4U 1746-37}?}
\shortauthors{Z.S. Li et al.}

\begin{document}

\title{An Ultra Low Mass and Small Radius Compact Object in \object{4U 1746-37}?}

\author{Zhaosheng Li \altaffilmark{1}}
\email{lizhaosheng@pku.edu.cn}

\author{Zhijie Qu \altaffilmark{1}}

\author{Li Chen \altaffilmark{2}}

\author{Yanjun Guo \altaffilmark{1}}

\author{Jinlu Qu \altaffilmark{3}}

\author{Renxin Xu \altaffilmark{1,4}}

\altaffiltext{1}{School of Physics and State Key Laboratory of Nuclear Physics and Technology, Peking University, Beijing 100871, China}
\altaffiltext{2}{Department of Astronomy, Beijing Normal University, Beijing 100875, China}
\altaffiltext{3}{Laboratory for Particle Astrophysics, Institute of High Energy Physics, CAS, Beijing 100049, China}
\altaffiltext{4}{Kavli Institute for Astronomy and Astrophysics, Peking University, Beijing 100871, P. R. China}

\begin{abstract}
Photospheric radius expansion (PRE) bursts have already been used to constrain the masses and radii of neutron stars. \textit{RXTE} observed three PRE bursts in \object{4U 1746-37}, all with low touchdown fluxes. We discuss here the possibility of low mass neutron star  in \object{4U 1746-37} because the Eddington luminosity depends on stellar mass. With typical values of hydrogen mass fraction and color correction factor, a Monte-Carlo simulation was applied to constrain the mass and radius of neutron star in \object{4U 1746-37}. \object{4U 1746-37} has a high inclination angle. Two geometric effects, the reflection of the far side accretion disc and the obscuration of the near side accretion disc have also been included in the mass and radius constraints of \object{4U 1746-37}. If the reflection of the far side accretion disc is accounted, a low mass compact object (mass of $0.41\pm0.14~M_{\odot}$ and radius of $8.73\pm1.54~\rm km$ at 68\% confidence) exists in \object{4U 1746-37}. If another effect
operated, \object{4U 1746-37} may contain an ultra low mass and small radius object ($M=0.21\pm0.06~M_{\odot},~R=6.26\pm0.99~\rm km$ at 68\% confidence). Combined all possibilities, the mass of \object{4U 1746-37} is $0.41^{+0.70}_{-0.30}~M_\odot$ at 99.7\% confidence. For such low mass NS, it could be reproduced by a self-bound compact star, i.e., quark star or quark-cluster star.

\end{abstract}


\section{Introduction}

The equation of state (EoS) of superdense matter is one of the key questions in astrophysics and nuclear physics. Neutron stars (NSs, hereafter, NS refers to all kinds of pulsar-like compact objects.) in the Universe provide us an unique opportunity to approach it. Generally, two categories of EoS were widely discussed, which can produce gravity-bound NS and self-bound NS \citep{Glendenning,Haensel07}, respectively. All of them proposed distinct mass-radius relations. The EoSs of self-bound NS predicted $M\propto R^3$ ($M$ and $R$ are the mass and radius of NS) for low mass NS. Moreover, the minimum mass of self-bound NS can reach as low as planet-mass \citep{Xu03,Horvath12}, while the low limit mass of gravity-bound NS is about $0.1~M_{\odot}$ (e.g., \citet{AP97,Glendenning99}). The measurements of the radius and mass of NS, as well as searching extremely low mass NS can provide useful information to test various theoretical EoSs.

The mass of NS can be precisely determined in double NS system or white dwarf--neutron star system (see \citet{Lattimer12} for all NSs with measured masses). Especially, \citet{Janssen08} found a very low mass NS ($<~1.17~M_{\odot}$ at 95.4\% confidence) in \object{PSR J1518+4904}, which might be the least massive compact object in double NS system. The direct measurement of the radius of NS, however, is still difficult. The measurement of NS radius is very critical for EoS constraining. \citet{Fortin14} claimed that the NSs with mass in the range $1.0-1.6~M_{\odot}$ should be larger than 12 km, otherwise, the presence of hyperons in neutron star cores are ruled out. And then, the so-called hyperon puzzle is arising (e.g., \citealt{Bednarek12}). Several methods were proposed to constrain the radius and mass of NS, such as fitting the thermal spectra from quiescent low-mass X-ray binaries (LMXBs) in globular clusters \citep{Guillot13}, simulating X-ray pulsar profiles \citep{Leahy04}, and
photospheric radius
expansions (PRE) bursts (see \citet{Bhat10} for a review).

Type I X-ray bursts in LMXBs are sudden energy release process, which last ten to hundred seconds and can emit as high as Eddington Luminosity ($\sim3.79\times10^{38}~\rm erg/s$). In classical view, type I X-ray bursts are powered by the unstable thermonuclear burning of H/He accreted on the neutron star surface through its companion star Roche-lobe overflowing. Most of the spectra of type I X-ray bursts can be well fitted by a pure black body spectrum. PRE burst, a special case of type I X-ray burst, were phenomenally distinguished from the time resolved spectra. At the touchdown moment, where the black body temperature and its normalization reach their local maximum and minimum during X-ray burst respectively, the referred bolometric luminosity corresponds to its Eddington luminosity, that is, the radiation pressure is balanced by the gravity. After the touchdown point, the residual thermal energy cool on the whole surface of neutron star during burst tail. So, the mass and radius of neutron star could be
constrained if the distance to the source was measured independently, i.e., in globular clusters \citep{Sztajno87,Ozel09}.

In the assumption of spherically symmetric emission, the Eddington luminosity is expressed as \citep{Lewin93},
\begin{equation}
 L_{\rm{Edd}}=\frac{8\pi Gm_{\rm p}Mc[1+(\alpha_{\rm{T}}T_{\rm{e}})^{0.86})]}{\sigma_{T}(1+X)(1+z(R))},
\end{equation}
where, $G$, $c$, $\sigma_{T}$ are the Gravitational constant, the speed of light and the Thompson scattering cross-section, respectively; $m_{\rm p}$ is the mass of the proton, $X$ is the atmosphere's hydrogen mass fraction ($X=1$ for pure hydrogen), $T_e$ is the effective temperature of NS atmosphere, $\alpha_{T}$ describes the temperature dependence of the electron scattering opacity. The factor $1+z(R)=(1-2GM/Rc^2)^{-1/2}$, is the gravitational redshift correction for strong gravity field on the surface of NS. \citet{kuulkers03} analyzed all PRE bursts in globular clusters with known distance, and discussed the potential advantage of PRE bursts as ``standard candle''. \citet{Galloway08} argued that the luminosity of PRE bursts were intrinsically affected by the mass and radius of NS, the variation of photosphere composition. Especially, two low luminosity sources during PRE bursts, \object{4U 1746-37} and \object{GRS 1747-312}, emitted too
faint to reach Eddington luminosity in the assumption of $1.4M_{\rm \odot}$. However, the possibility of the observed low flux due to the existence of low mass NS, i.e., $0.7M_{\rm \odot}$ \citep{Sztajno87}, cannot be ruled out.

We interpret that a low mass NS inside \object{4U 1746-37} and \object{GRS 1747-312} can explain their low touchdown fluxes in PRE bursts. However, a peculiar X-burst from \object{GRS 1747-312} exhibited significant variation of apparent radius in the cooling tail \citep{Zand03}. The color correction factor as well as emission area may simultaneously change similar as the case in \object{4U 1820-30} \citep{Garcia13}. In this work, we only discuss the possibility of a low mass NS in \object{4U 1746-37}.

Compared with very early works by \citet{Sztajno87}, we consider the touchdown fluxes, instead of peak fluxes, observed by \textit{RXTE} in \object{4U 1820-30} as its Eddington flux. Moreover, the reflection or obscuration by accretion disc are accounted separately \citep{Galloway08b}. The accretion rate enhancement during X-ray bursts are checked \citep{Worpel13,Zand13}. The effects of extremely extended photosphere at touchdown moment are also investigated \citep{Steiner10}.

In Section 2, the \textit{RXTE} observations of \object{4U 1746-37} will be briefly presented. In Section 3, we introduce the mass-radius constraints of \object{4U 1746-37}. We give the results and discussions in Section 4 and 5.

\section{\textit{RXTE} Observations}
During its 15 years operation, \textit{RXTE} observed over 1000 X-ray bursts, which were detailedly analyzed in \citet{Galloway08}. The high quality data provided an opportunity to research the time resolved spectra of X-ray bursts. The PRE bursts, a special type of X-ray bursts, emitted Eddington luminosity and cooled on the whole surface of NS with small uncertainties \citep{Guver12a,Guver12b}, which were utilized to determine the $M$ and $R$ of NS \citep{Ozel09,Guver10a,Guver10b,Ozel12}. The dominant uncertainties of $M$ and $R$ originated from the error of the distance to source \citep{Sala12}.

The touchdown fluxes and black body normalizations ($A$) were obtained in the time resolved spectra of PRE bursts. When extracting the time resolved spectra, several assumptions were made first \citep{Worpel13}. The spectra of persistent emission during bursts were stable and invariant. The net contribution of a burst was archived by subtracting its pre-burst intensity, which arose from accretion. \citet{Zand13} observed a type I X-ray burst in \object{SAX J1808.4-3658} with \textit{RXTE} and \textit{Chandra} simultaneously, and found obvious excess of low and high energy photons when fitted the burst spectrum with black body. \citet{Worpel13} explained that the excesses at low and high energies in \object{SAX J1808.4-3658} and other PRE bursts were due to accretion enhancement during the burst, analogous to Poynting-Robertson drag effect. \citet{Zand13} introduced ``$f_a$'' model to account for the contribution of persistent emission. \citet{Worpel13} found most of the spectra the factor $f_a$ were
significantly larger than unity, especially for \object{SAX J1808.4-3658} ($f_a=17.75$). We check this kind of accretion rate enhancement during type I X-ray burst in \object{4U 1746-37}. 

\subsection{Data Reduction}
\object{4U 1746-37} is a low mass X-ray binary (LMXB) located in the Globular Cluster \object{NGC 6441}. The distance to \object{NGC 6441} is $11.0^{+0.9}_{-0.8} ~\rm kpc$ \citep{kuulkers03}. From the type I X-ray burst catalog of \textit{RXTE} \citep{Galloway08}, three PRE bursts were identified in \object{4U 1746-37} (observation ID: 30701-11-03-000, 30701-11-04-00, 60044-02-01-03, hereafter, we cited as Burst I, II, III, respectively). In order to check the accretion rate enhancement consequence during type I X-ray burst \citep{Worpel13}, we re-analyzed these three PRE bursts of \object{4U 1746-37}, which were collected by the Proportional Counter Array (PCA) on board of \textit{RXTE}.  The time resolved spectra were extracted from appropriate model files (science event or Good Xenon), which covered the whole burst interval in the energy 2-60 keV. The dead time corrections were made following the process suggested by the \textit{RXTE} team\footnote{http://heasarc.nasa.gov/docs/xte/recipes/pca\_deadtime.
html}. We fitted the spectra in the range 3-22 keV, and added 0.5\% systematic error. We fixed the hydrogen column density at $0.26\times10^{22}~ \rm cm^{-2}$ obtained from \textit{BeppoSAX} \citep{Sidoli01}, which has higher sensitivity at low X-ray energy than PCA/\textit{RXTE}. The deadtime correction factor ranges of each observation are listed in Table~\ref{tb-1}.

\subsection{Persistent Emission}
\begin{figure}
\includegraphics[width=5 cm,angle=270]{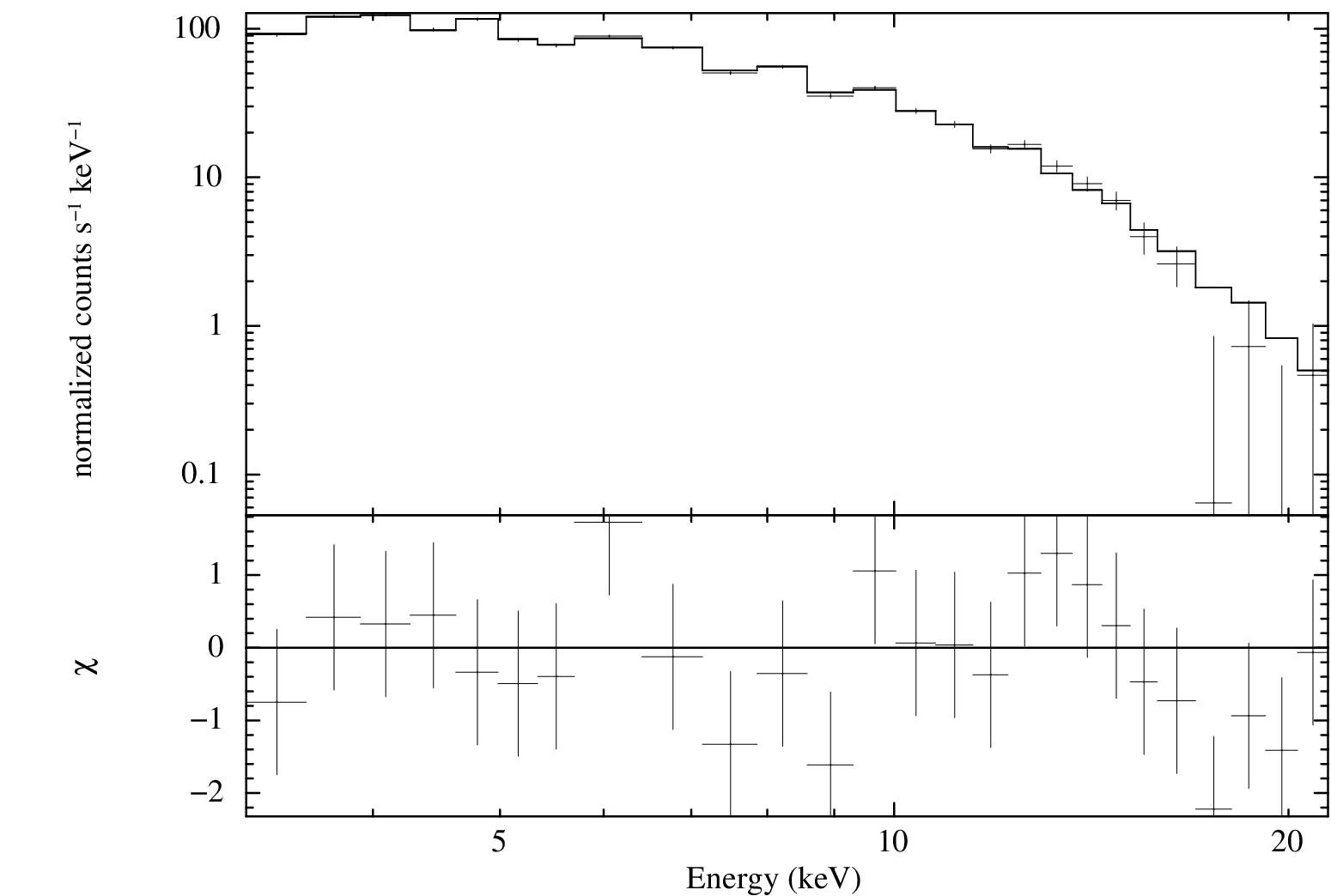}
\caption{The persistent emission spectrum of 4U 1746-37 (observation ID: 30701-11-03-000). The reduced-$\chi^2$ is 1.01, which implies a good fitting to the data.}
\label{fig:per}
\end{figure}

For each PRE burst in \object{4U 1746-37}, a 16 s interval prior to the trigger moment was regarded as persistent emission, which contains emission from the source as well as background from the instrument. We utilized the ``bright'' source model ($>$ 40 counts/s/PCU) to estimate the instrumental background with \textbf{runpcabackest} procedure. The persistent emission can be well fitted by absorbed black body plus power law (\textbf{wabs(bbodyrad+powerlaw)} in Xspec). Fig.~\ref{fig:per} shows the fit to the persistent spectrum and the residuals of Burst I. The reduced $\chi^2$ is 1.01 for 22 degree of freedom. For the other two PRE bursts, the reduced $\chi^2$ are 0.98 (Burst II) and 0.82 (Burst III), indicating well fitting to the data.
\subsection{Fitting the Burst Spectra}
The net burst spectrum can be represented by a pure black body with the interstellar absorption. \citet{Worpel13} introduced a $f_a$-model to account for the variation of persistent emission amplitude, which presents as
\begin{equation}
 S(E)=A(E)\times B(E; T_{\rm BB}, A_{\rm BB})+f_a \times P(E)-b(E)_{\rm inst},
 \label{equ:spec}
\end{equation}
here, $A(E)$ is the absorption correction, $B(E; T_{\rm BB}, A_{\rm BB})$ is the black body spectrum with temperature $T_{\rm BB}$ and normalization $A_{\rm BB}$, $P(E)$ is the persistent emission and $b(E)_{\rm inst}$ is the instrumental background. The parameter $f_a$ accounts for the contribution from the persistent emission, i.e., $f_a=1$ means that the amplitude of persistent emission is exactly the same as the moment before X-ray burst trigger. Note that the $f_a$ model is applied, with assuming that only the amplitude of persistent emission can change. The $f_a$ distribution of type I X-ray burst from \citet{Galloway08} catalog is peaked at 1, and biased towards higher values \citep{Worpel13}. It implied that the accretion rate increases during X-ray burst analogous to the Poynting-Robertson effect.

\begin{figure}
 \plotone{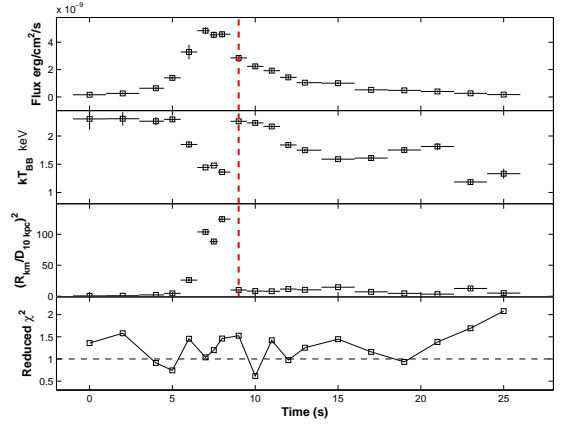}
 \caption{The time resolved spectra of PRE burst in 4U 1746-37 (observation ID: 30701-11-03-000). The red dashed line labels the touchdown moment. The 1-$\sigma$ errors are displayed. For some data, the errors are smaller than the symbols.}
 \label{fig:03-000}
\end{figure}

\begin{figure}
 \plotone{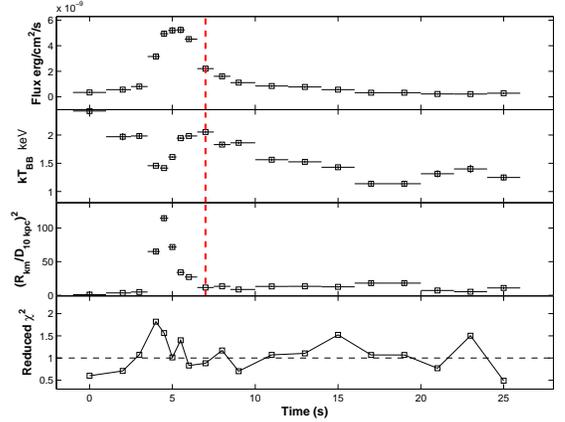}
 \caption{The time resolved spectra of PRE burst in 4U 1746-37 (observation ID: 30701-11-04-00).}
 \label{fig:04-00}
\end{figure}

\begin{figure}
  \plotone{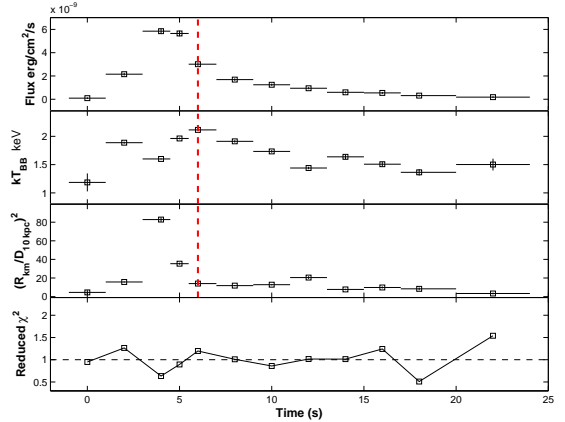}
 \caption{The time resolved spectra of PRE burst in 4U 1746-37 (observation ID: 60044-02-01-03).}
 \label{fig:01-03}
\end{figure}

We also attempted to find whether the persistent emission varied or not in \object{4U 1746-37}.  When the $f_a$ model was used, we applied the $f$-test to check the requirement of adding this extra parameter. We found that the $f_a$ model cannot produce distinctly better reduced $\chi^2$. It implies that even if the accretion rate increased during type I X-ray burst in \object{4U 1746-37}, its contribution to the burst spectrum can be neglected. We generated the time resolved spectra of three PRE bursts in Fig.~\ref{fig:03-000}, Fig.~\ref{fig:04-00} and Fig.~\ref{fig:01-03}. The bolometric flux, the black body temperature, the black body normalization and the reduced $\chi^2$ are showed. The bolometric flux was calculated from equation (3) in \citet{Galloway08}.
The error of bolometric flux was estimated from the uncertainty propagation. All quoted errors are at 68\% confidence level.

In Fig.\ref{fig:03-000}, the reduced $\chi^2$ on the cooling tail are relative large compared with the expansion phase and contraction phase, so does the $f_a$ model. The touchdown fluxes are $(2.86\pm0.16) \times 10^{-9}~ \rm erg/s/cm^2$, $(2.21\pm0.14) \times 10^{-9}~ \rm erg/
s/cm^2$ and $(3.01\pm0.13) \times 10^{-9}~ \rm erg/cm^2/s$. The corresponding peak fluxes are $(4.84\pm0.25) \times 10^{-9}~ \rm erg/s/cm^2$, $(5.23\pm0.26)\times 10^{-9}~ \rm erg/s/cm^2$ and $(5.84\pm0.23)\times 10^{-9}~ \rm erg/s/cm^2$. Meanwhile, the factor $F_{\rm p}/F_{\rm TD}$ is $2.0\pm0.3$. If the cooling tails were truncated at $0.5\times10^{-9}~\rm erg/cm^2/s$, we obtained the apparent area during the cooling tail $10.9\pm4.2~{\rm(km/10~kpc)}^2$, which is smaller than $15.7\pm 2.4 ~\rm (km/10~kpc)^2$  provided by \citep{Guver12a}, but with a larger error. Since, we only used three PRE bursts, and  did not group the apparent areas as a function of flux.

\begin{figure}
 \plotone{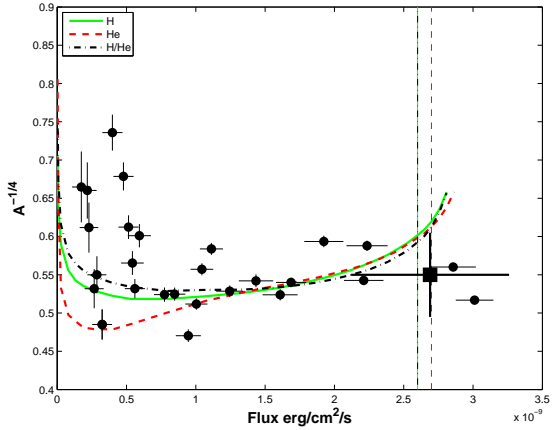}
 \caption{The evolution curve of $A^{-1/4}$. \citet{Suleimanov11} suggested that it reflects the variation of the color correction factor during the cooling tail. Three theoretical models for $\log g=14$ ($g$ is the gravitational constant) and $R=14.8~\rm km$ are displayed by solid green curve (pure Hydrogen), red dashed curve (pure Helium) and black dash-dotted curve (solar mixture of H/He with metal abundance of $Z=0.01Z_{\odot}$) \citep{Suleimanov12}, which are able to fit the data at high flux (i.e. larger than $0.5\times10^{-9}~\rm erg/cm^2/s$). However, the models can not fit the data well at low flux. The vertical lines mark the Eddington fluxes for aforementioned theoretical models prediction displaying with the same line styles. Note that these theoretical curves are calculated for a particular mass and radius of NS. The black square shows the 1-sigma confidence interval of the Eddington flux and cooling area measured from the touchdown method.}
 \label{fig:color}
\end{figure}

\citet{Suleimanov11} proposed that the color correction factor are apparently changed when the luminosity close to its Eddington limit. In Fig.~\ref{fig:color}, the $A^{-1/4}-Flux$ correlation is shown and fitted by three theoretical models \citep{Suleimanov12}. The data are well fitted at high flux ($F_{\rm TD}=2.65\times10^{-9}~ \rm erg/cm^2/s$ and $[R(1+z)/D_{10}]^{-1/2}=0.35$ for pure H, $F_{\rm TD}=2.7\times10^{-9} ~\rm erg/cm^2/s$ and $[R(1+z)/D_{10}]^{-1/2}=0.36$ for pure He, $F_{\rm TD}=2.65\times10^{-9}~ \rm erg/cm^2/s$ and $[R(1+z)/D_{10}]^{-1/2}=0.37$ for mixture of H/He). At low flux, the data deviate from the prediction of models, which also appears in \object{GS 1826-24} \citep{Zamfir12}.
\citet{Guver12a} calculated $f_c$ for different X-ray burst atmosphere models and concluded that $f_c$ is weakly dependent on the temperature if the black body temperature is less than 2.5 keV. From the time resolved spectra of \object{4U 1746-37}, the black body temperature are all in $1-2$ keV. Hence, the color correction factor is chosen as $1.3-1.4$ to account for the different theoretical model predictions.

The standard deviations of $F_{\rm TD}$ and $A$ contain three parts, the observed errors ($\sigma_{F_{\rm TD,~obs}},~\sigma_{A_{\rm obs}}$), the systematic errors ($\sigma_{F_{\rm TD,~sys}},~\sigma_{A_{\rm sys}}$) and the absolute calibration errors ($\sigma_{F_{\rm TD,~cal}},~\sigma_{A_{\rm cal}}$), which are
\begin{equation}
 \sigma^2_{F_{\rm TD}}=\sigma^2_{F_{\rm TD,~obs}}+\sigma^2_{F_{\rm TD,~sys}}+\sigma^2_{F_{\rm TD,~cal}},
 \label{equ:flux_err}
\end{equation}
and
\begin{equation}
 \sigma^2_A=\sigma^2_{A_{\rm obs}}+\sigma^2_{A_{\rm sys}}+\sigma^2_{A_{\rm cal}},
 \label{equ:area_err}
\end{equation}
if these errors are independent with each other. Here, the 10\%  absolute calibration errors are applied \citep{Tsujimoto11}. Since the systematic errors were $3\%-8\%$ for apparent radii \citep{Guver12a} and $\sim10\%$ for touchdown fluxes \citep{Guver12b}. We adopted 8\% and 10\% systematic errors for apparent radius and touchdown flux, respectively.   So, the mean touchdown flux and apparent area are $(2.69\pm0.57)\times 10^{-9} ~\rm erg/cm^2/s $ and $10.9 \pm 4.4~\rm (km/10~kpc)^2$ for these PRE bursts. We note that two PRE bursts were observed by \textit{EXOSAT} with peak fluxes $(1.0\pm0.1)\times10^{-8}~\rm erg/cm^2/s$ and touchdown fluxes about $2.2-4.2 \times 10^{-9}~\rm erg/cm^2/s$ \citep{Sztajno87}. The touchdown flux observations of \textit{RXTE} and \textit{EXOSAT} for \object{4U 1746-37} were consistent with each other.
It should be mentioned that \citet{Sztajno87} treated the peak flux as the Eddington flux. Here, we adopted the touchdown flux as its Eddington flux as suggestion in \citet{Ozel09}.

\object{4U 1746-37} has a high system inclination angle ($i\sim90^\circ$). In such systems, the touchdown fluxes were systematically smaller than the peak fluxes. \citet{Galloway08b} found that the ratios between the peak flux ($F_{\rm p}$) and the touchdown flux ($F_{\rm TD}$) are larger than $\sim 1.6$ in dipping binaries. They discussed two geometric interpretations of this ratio, the reflection of the far side accretion disc and the obscuration of the near side accretion disc. For the first scenario, the different between $F_{\rm p}$  and $F_{\rm TD}$ is due to the extra contribution from the far side disc reflection at the peak flux moment. So, the touchdown flux exactly corresponds to its Eddington flux. For the second scenario, it is the anisotropies of persistent and burst emission, which have been discussed for a long time \citep{Lapidus85,Fujimoto88,Zamfir12}. If the geometrically thin accretion disc extends close to the neutron star surface, it will intercept $\sim 1/4$
of the burst radiation, and re-radiate along the disc axis \citep{Lapidus85}. \citet{Fujimoto88} introduced an anisotropy parameter $\xi$, and expressed the actual luminosity of burst emission as $L= {4\pi} D^2\xi F_{\rm b}$, where $F_{\rm b}$ is the observed burst flux. \citet{Lapidus85} suggested the approximate approach for $\xi$,

\begin{equation}
 \xi^{-1}=\frac{1}{2}+|\cos{i}|.
\end{equation}
In an edge on binary system, the anisotropy parameter $\xi$ is 2.  Moreover, it can be estimated as $F_{\rm p}/F_{\rm TD}$. Since, the obscured fraction of burst emission at the peak flux moment is much smaller than the one at the touchdown moment \citep{Galloway08b}. Hence, at this circumstance, the touchdown flux as well as emission area in the cooling tail should be corrected to larger values with the factor $F_{\rm p}/F_{\rm TD}$. Here, $F_{\rm p}/F_{\rm TD}$ is $2.0\pm0.3$ for 4U 1746-37, which is consistent with the above mentioned prediction. We considered these two geometric effects separately.

\section{The constraining of $M$ and $R$}
In PRE bursts, the mass and radius of NS are constrained from the relations \citep{Ozel09},
\begin{equation}
 F_{\rm TD}=\frac{GMc}{k_{\rm es}D^2}(1-\frac{2GM}{Rc^2})^{1/2},
 \label{equ:flux}
\end{equation}
and
\begin{equation}
 A=\frac{R^2}{D^2f^4_{\rm c}}(1-\frac{2GM}{Rc^2})^{-1},
 \label{equ:area}
\end{equation}
where, $k_{\rm es}=0.2(1+X)~\rm cm^2/g$ is opacity to electron scattering, $f_{\rm c}$ is color correction factor. In order to constrain the mass and radius of NS properly, the uncertainties of photosphere composition ($X$), distance, and color correction factor should be taken into account together. \citet{Ozel09} proposed a Bayesian framework to estimate the mass and radius of NS. They set each quantity with independent probability distribution functions, and then, the joint probability density of mass and radius is expressed as,
\begin{align}
 P(D, X, f_{\rm c}, M, R)=\frac{1}{2}|J(\frac{F_{\rm TD}, A}{M, R})|P(D)P(X)\nonumber \\
 P(f_{\rm c})P(F_{\rm TD})P(A)dDdXdf_{\rm c}dMdR,
 \label{equ:prob}
\end{align}
here, the Jacobian of the transformation from the pair $(F_{\rm TD}, A)$ to $(M, R)$ is supposed to be
\begin{align}
 J(\frac{F_{\rm TD}, A}{M, R})=\frac{2GcR}{k_{\rm es}D^4 f_{c}^{4}}(1 - 4\frac{GM}{Rc^2})(1-\frac{2GM}{Rc^2})^{-3/2}.
 \label{equ:jacob}
\end{align}
\citet{Ozel12} made a correction for this expression compared to Equation (\ref{equ:jacob}) in \citet{Ozel09}, but a factor of 2 is still missing. Although, the mass-radius confident regions are not effected by the constant factor in Equation (\ref{equ:prob}) when the joint probability density is normalized. Integrated Equation (\ref{equ:prob}) over distance, the joint probability distribution of $M$ and $R$ is obtained.

In this work, a Monte Carlo method is applied to constrain $M$ and $R$ of NS, which shows high efficiency \citep{Li12}. We produce two series of simulated $F'_{\rm TD}$ and $A'$, which satisfy $F'_{\rm TD}\sim N(F_{\rm TD,~ obs},\sigma^2_{F_{\rm TD}})$ and $A'\sim N(A_{\rm obs},\sigma^2_{A})$, respectively. Here, $N(F_{\rm TD,~ obs},\sigma^2_{F_{\rm TD}})$ denotes that $F'_{\rm TD}$ is normally distributed random values with expectation $F_{\rm TD,~ obs}$ and standard deviation $\sigma_{F_{\rm TD}}$.   $N(A_{\rm obs},\sigma^2_{A})$ has a similar definition. We also assign flat distributions for $X$, $f_{\rm c}$, which are correspondingly represented as $X'\sim U[X-{\rm d}X,~X+{\rm d}X],~ f'_{\rm c}\sim U[f_{\rm c}-{\rm d}f_{\rm c},~f_{\rm c}+{\rm d}f_{\rm c}]$. Especially, the distance to the source has asymmetric errors. In order to simplify the simulation, we adopt $D'\sim\{N(D_0,\sigma^2_{D_1})1_{[D_0,\infty)}(D)+N(D_0,\sigma^2_{D_2})1_{(-\infty,D_0)}(D)\}$, where $1_{[D_0,\infty)}(D)$ denotes the
indicator function of set $[D_0,\infty)$, $D_0=11~\rm kpc$, $\sigma_{D_1}=0.9~\rm kpc$ and $\sigma_{D_2}=0.8~\rm kpc$\footnote{A flat distribution of $D$ is also attempted. The $M-R$ confidence contours are shifted negligibly. }. The hydrogen mass fraction and the color correction factor are set as $0.35\pm0.35$,  $1.35\pm 0.05$ and \citep{Suleimanov11, Guver12a}, respectively.  For each
pair of $(F'_{\rm TD},~A', ~D', ~f'_{\rm c},~ X')$, the $M$ and $R$ of NS are solved from Equation (\ref{equ:flux}) and (\ref{equ:area}), if the solutions exist. For a certain large samples (i.e., $10^7$), the confidence regions of $M$ and $R$ are obtained.

\section{Results}
\begin{figure*}\centering
\includegraphics[height=6.5 cm]{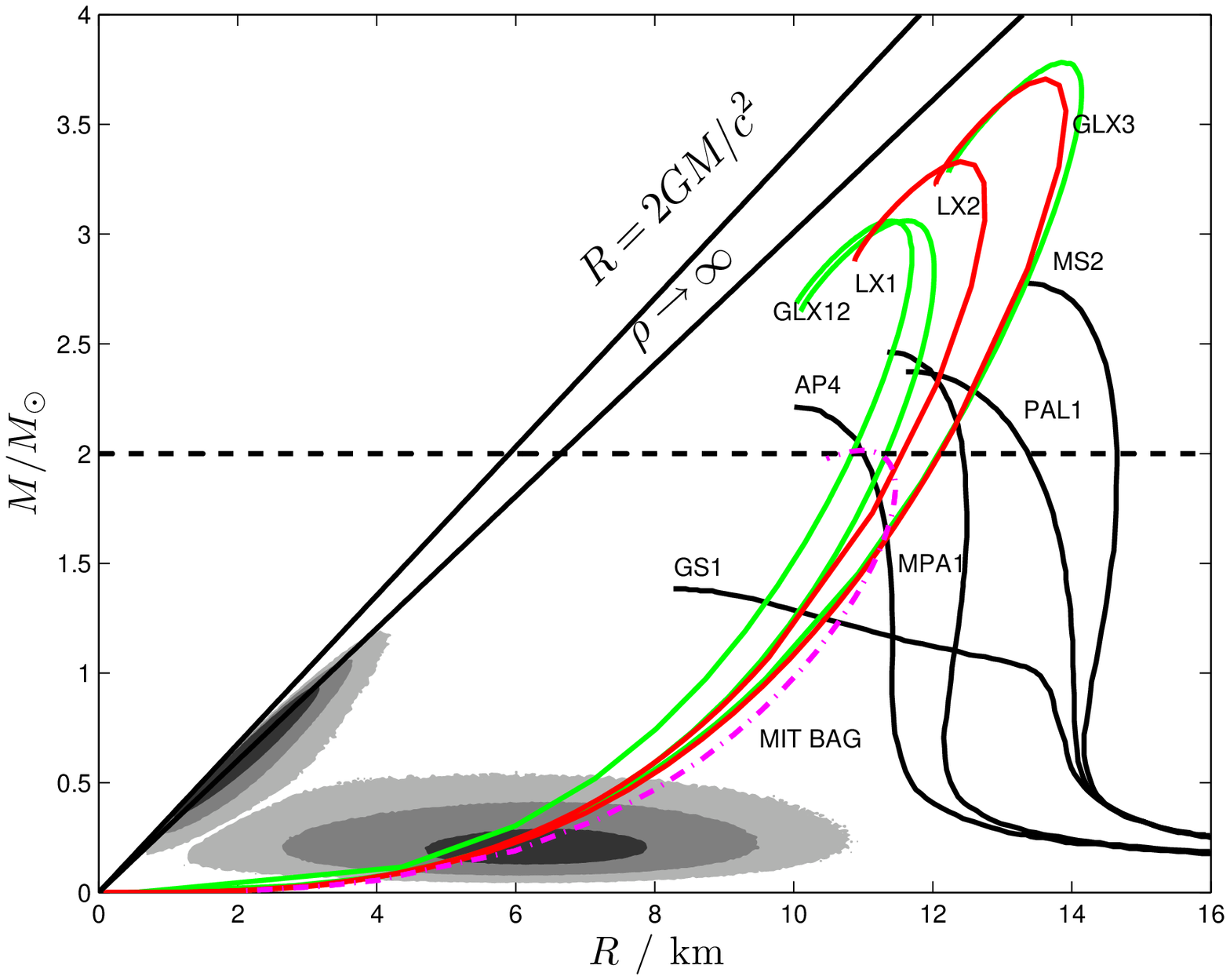}
\includegraphics[height=6.5 cm]{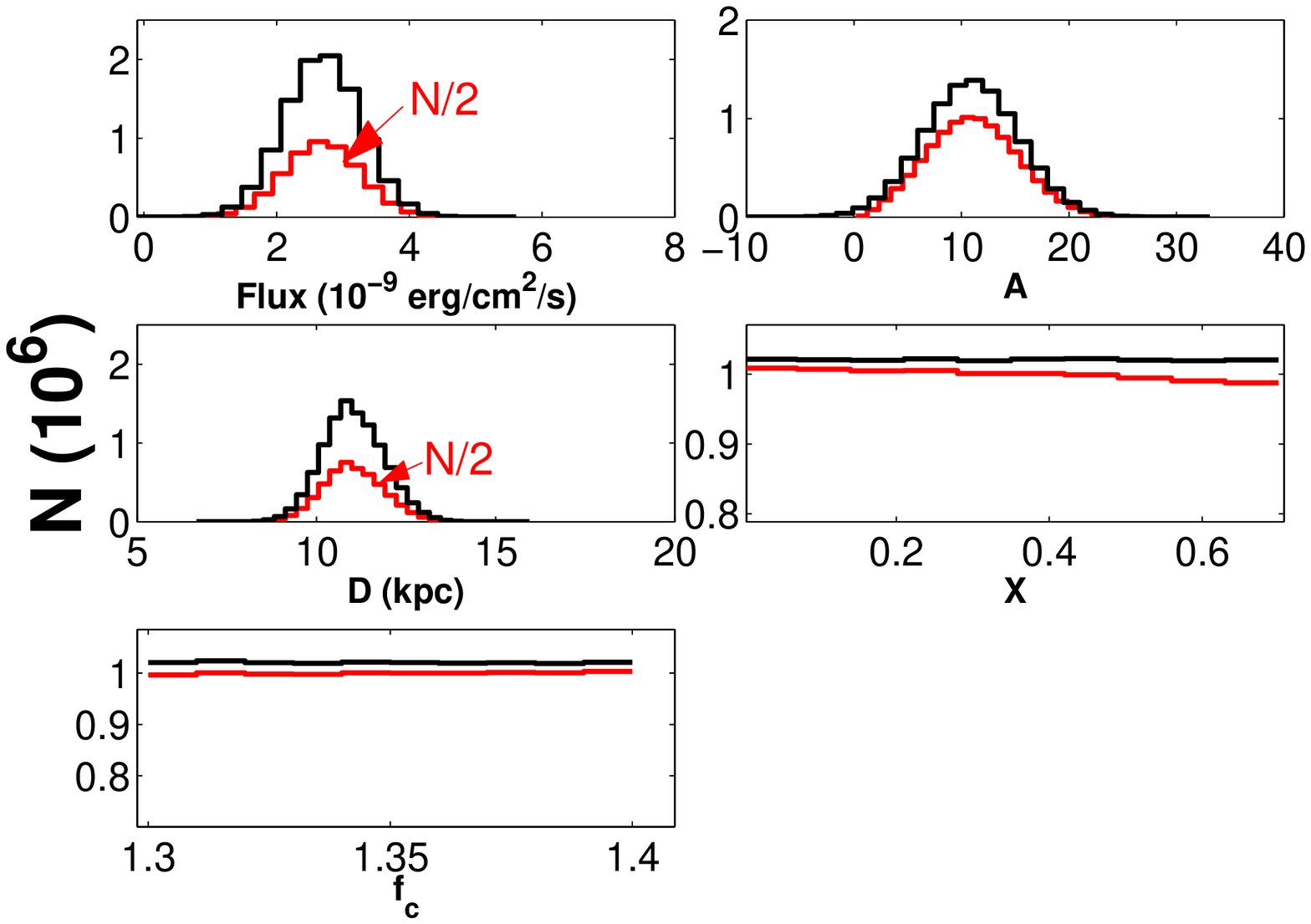}
\caption{\textit{Left panel}: the 1-, 2-, 3-sigma $M-R$ confidence regions of 4U 1746-37, which are based on the assumption that the touchdown flux corresponded to the Eddington flux.
The dashed line denotes two observed near $2M_\odot$ NSs. The left black lines show the the general relatively (GR) limit and the central density limit, respectively. Theoretical mass-radius relations for several NS EoS models are displayed, which were introduced by GS1 \citep{Glendenning99}, AP4 \citep{AP97}, MPA1 \citep{MPA87}, PAL1 \citep{PAL88}, MS2 \citep{MS96}, GLX123 \citep{Guo13}, LX12 \citep{Lai09,Lai13}. The purple dash-dotted line represents the bare strange stars obtained from MIT bag model EoS. In order to reach $M_{\rm max}=2M_{\odot}$, the bag constant equals to $57~{\rm MeV/fm^3}$. The first five gravity-bound NSs describe the same as in \citet{Lattimer07}. \textit{Right panel}: the prior (black lines) and posterior (red lines) distributions of all relative parameters. In order to show the flux and distance distributions clearly, the total numbers of posterior distributions in both subgraphs are divided by a factor of 2, because the prior and posterior distributions are quite similar.
 The simulation contains $10^7$ samples.}
\label{fig:dist1}
\end{figure*}

\begin{figure*}\centering
\includegraphics[height=6.5 cm]{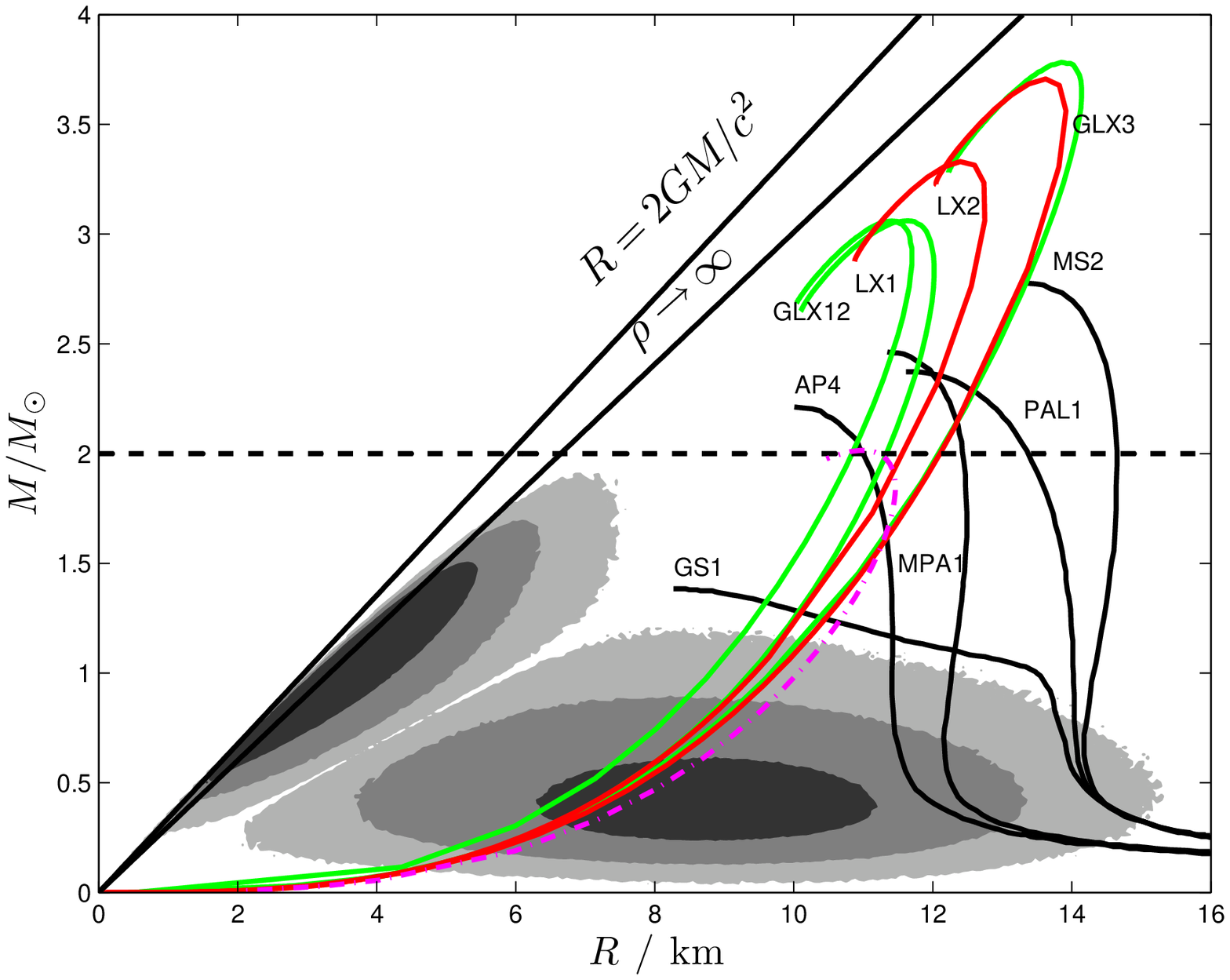}
\includegraphics[height=6.5 cm]{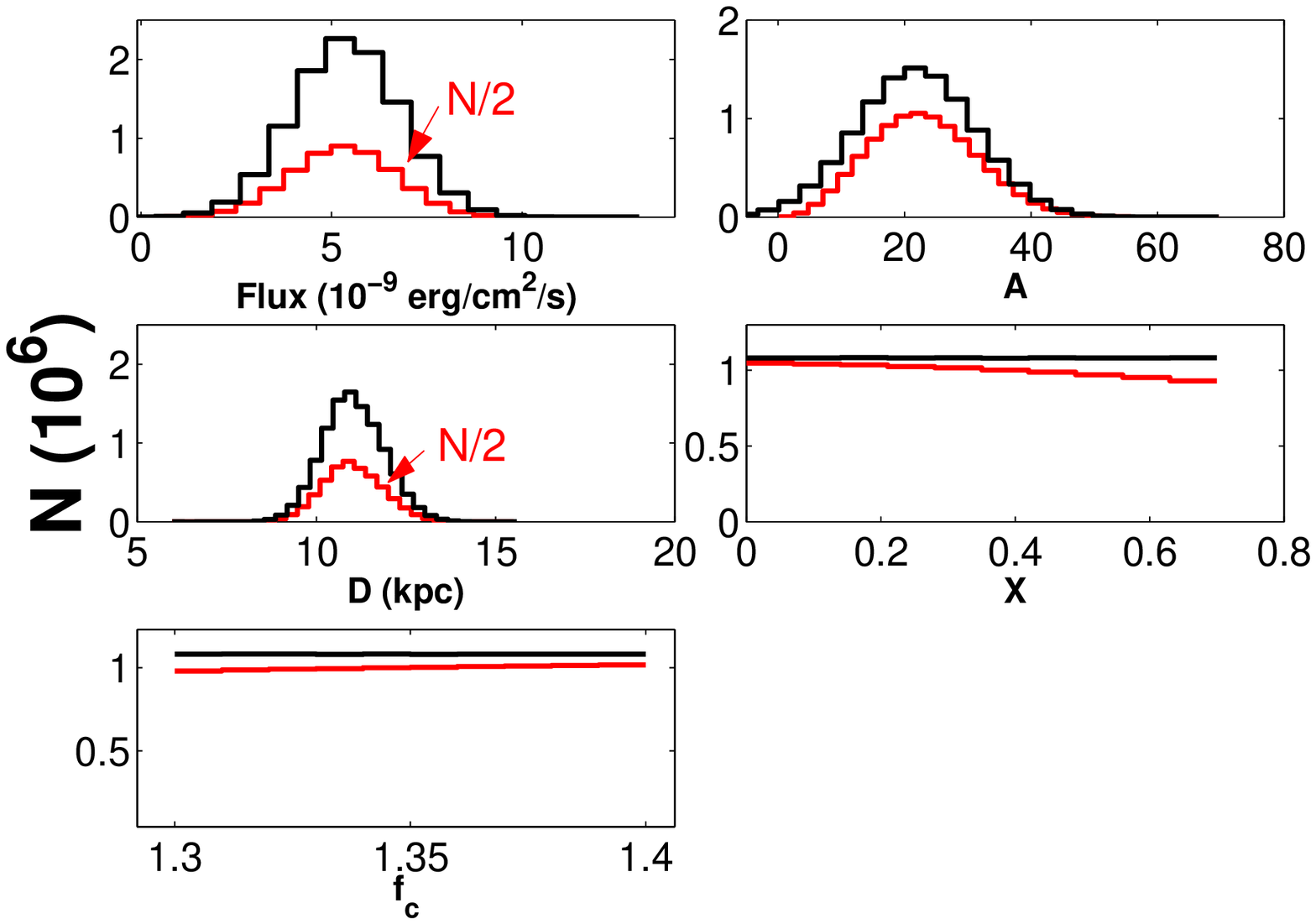}
\caption{Same as Fig.\ref{fig:dist1}. But, it based on the assumption that the touchdown flux as well as emission area were partially obscured by the accretion disc. For 4U 1746-37, the obscuration factor $F_{\rm p}/F_{\rm TD}$ is $2.0\pm0.3$ and its error is accounted in the  contours of $M-R$. }
\label{fig:dist2}
\end{figure*}

We applied a Monto-Carlo simulation to constrain the mass and radius of NS in \object{4U 1746-37}. The typical distributions of color correction factor and hydrogen mass fraction were utilized. The results are shown in Fig.~\ref{fig:dist1} and Fig.~\ref{fig:dist2}. The left panel in Fig.~\ref{fig:dist1} displays the 1-, 2-, 3-sigma confidence regions of the mass and radius of \object{4U 1746-37}, if the touchdown flux exactly corresponds to the Eddington flux. That is, the peak flux contained a significant fraction component from the reflection of the far side disc. If the accretion disc obscured a portion of emission area at the touchdown moment and in the cooling tail, $F_{\rm TD}$ and $A$ should be corrected with the factor $F_{\rm p}/F_{\rm TD}$, here, the factor $2.0\pm0.3$ was adopted. The confidence regions are displayed in the left panel of Fig.~\ref{fig:dist2}.  Ten EoSs are also plotted. It should be mentioned that in each case two regions are preferred. In Fig.~\ref{fig:dist1}, the mass and radius
of NS are $0.63\pm0.18~M_{\odot}$ and $2.14\pm0.61~\rm km$ for the
up-left part, or $0.21\pm0.06~M_{\odot}$ and $6.26\pm0.99~\rm km$ for the bottom-right part. In Fig.~\ref{fig:dist2}, the mass and radius of NS are $0.99\pm0.29~M_{\odot}$ and $3.55\pm1.14~\rm km$ for the up-left part, or $0.41\pm0.14~M_{\odot}$ and $8.73\pm1.54~\rm km$ for the bottom-right part.

We also checked the prior and posterior distributions of all related parameters. From the Fig.~\ref{fig:dist1} and Fig.~\ref{fig:dist2}, the posterior distributions are well consistent with prior ones.

The left contours cannot be reproduced by any EoS, because the mean densities of NS are much larger than the nuclear matter saturation density, and they are close the Schwarzschild radius. The results show that \object{4U 1746-37} contains a very low mass NS in the range $0.21-0.41~M_{\odot}$. If only the reflection of the far side disc effect existed, the touchdown flux equals to its Eddington flux. And then, we conclude a ultra low mass and small radius NS inside \object{4U 1746-37}.

\begin{figure*}\centering
\includegraphics[height=6.5 cm]{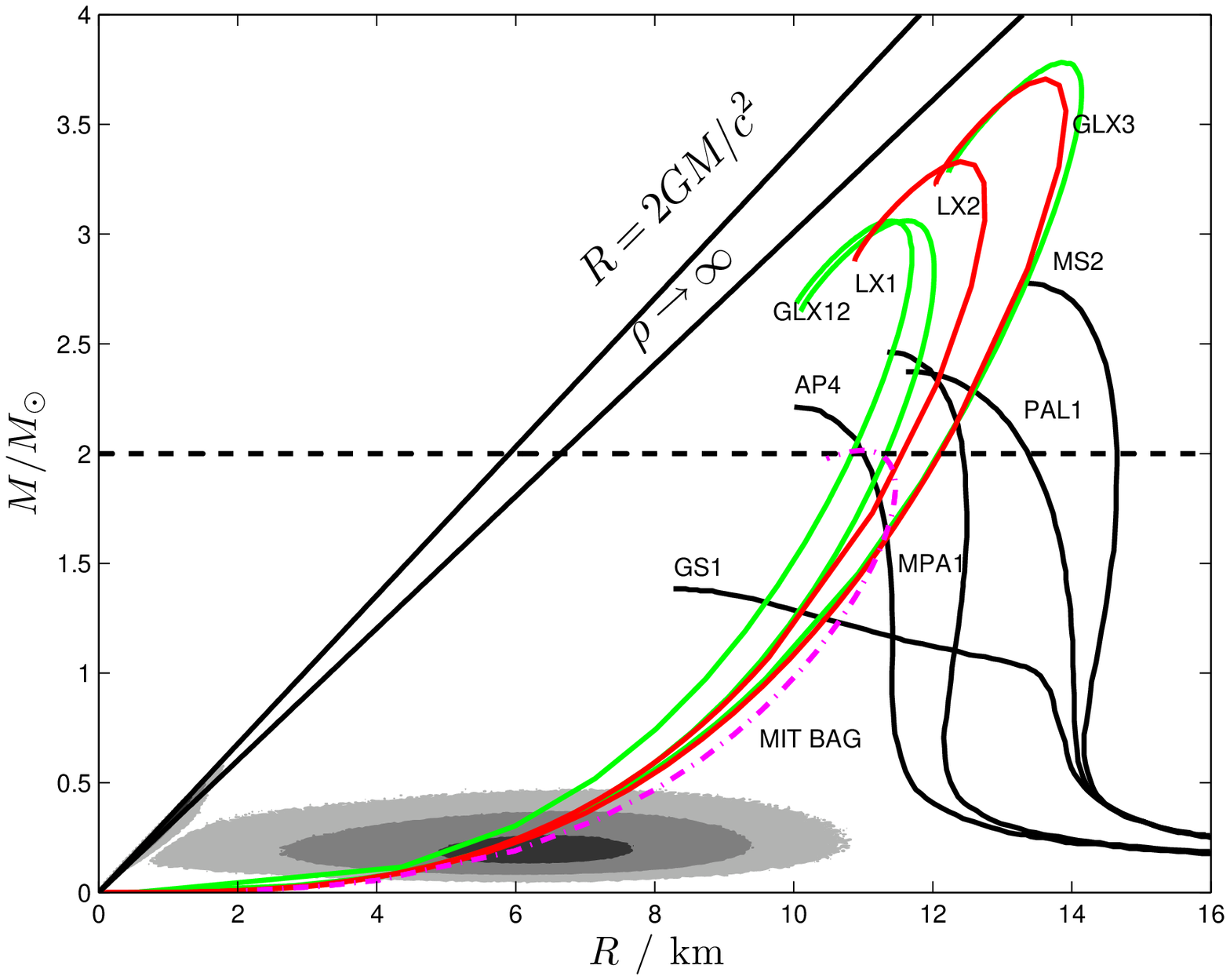}
\includegraphics[height=6.5 cm]{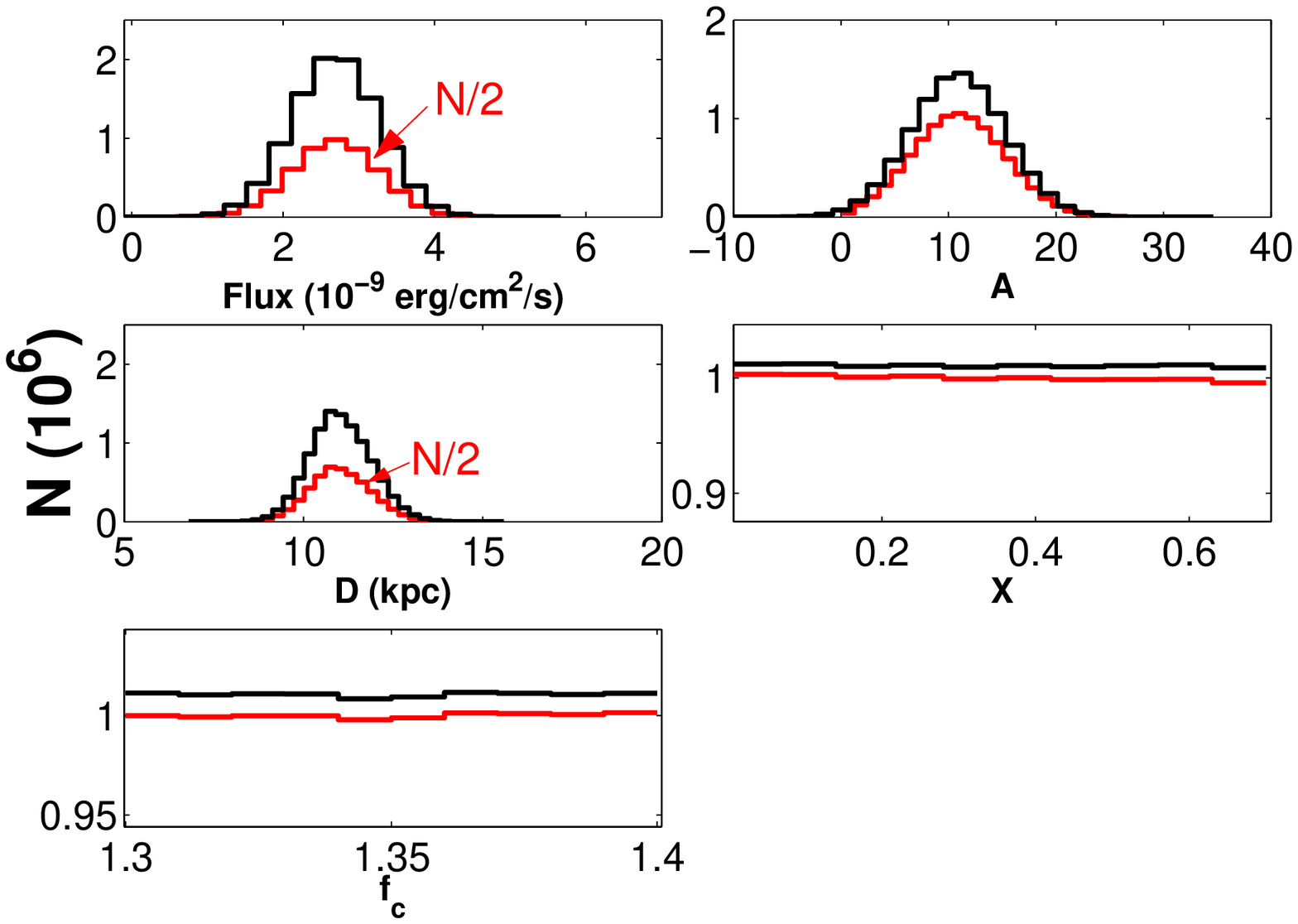}
\caption{Same geometric effect as Fig.\ref{fig:dist1}. The radius of photosphere at the touchdown moment is much larger than $R$. }
\label{fig:dist3}
\end{figure*}

\begin{figure*}\centering
\includegraphics[height=6.5 cm]{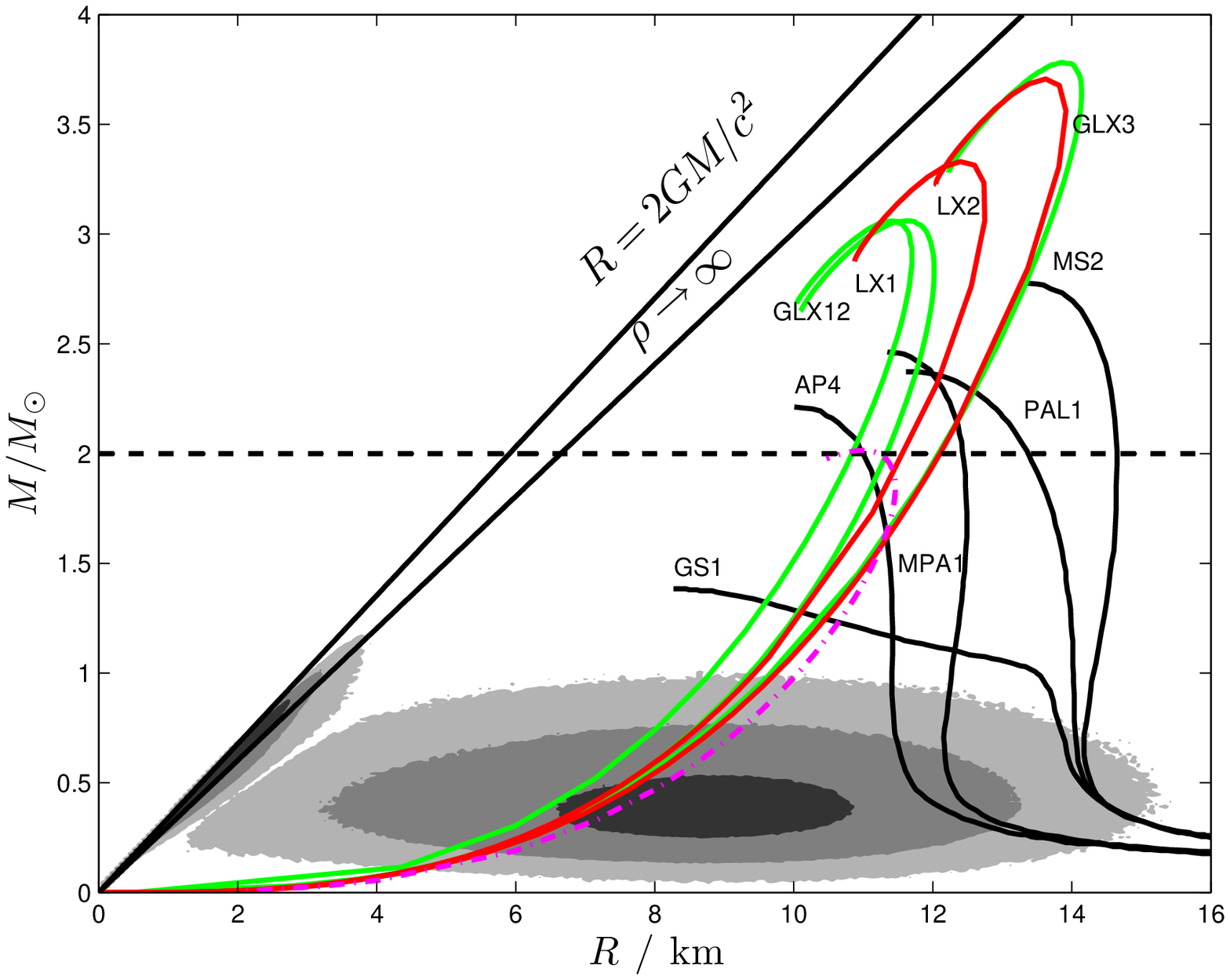}
\includegraphics[height=6.5 cm]{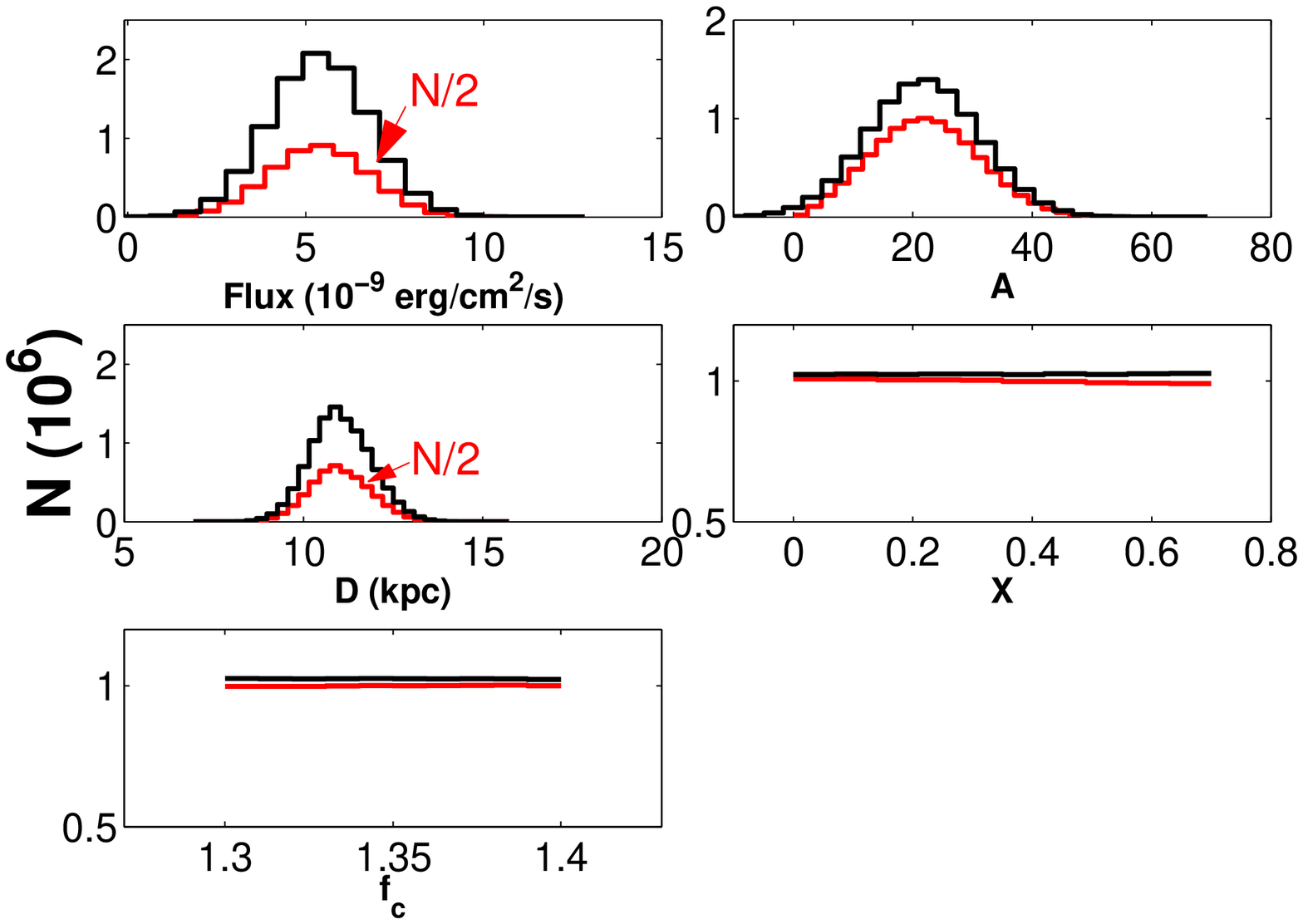}
\caption{Same geometric effect as Fig.\ref{fig:dist2}. It based on the assumption that the touchdown flux as well as emission area were partially obscured by the accretion disc, and the photosphere is extremely extended at the touchdown moment.}
\label{fig:dist4}
\end{figure*}

\citet{Steiner10} proposed that the photosphere could be still extended at the touchdown moment. At the extreme case, the radius of photosphere radius is much larger than the radius of NS, and then, the Eddington flux in Equation (\ref{equ:flux}) is reduced to
\begin{equation}
 F_{\rm TD}=\frac{GMc}{k_{\rm es}D^2},
 \label{equ:flux_1}
\end{equation}
the expression of apparent area in Equation (\ref{equ:area}) remains unchanged. The simulation results are shown in Fig.~\ref{fig:dist3} and Fig.~\ref{fig:dist4}. Each mass of NS corresponds to two different radius solutions. Compared with Fig.~\ref{fig:dist1} and Fig.~\ref{fig:dist2}, the left contours are shrunk in Fig.~\ref{fig:dist3} and Fig.~\ref{fig:dist4} and the right contours are shifted negligibly.

\section{Discussion and Conclusions}
Plenty of theoretical NS EoSs were proposed. Hadron star and hybrid/mixed star are gravity-bound, which are covered by crusts with nuclei and electrons, whereas quark star and quark-cluster star are strongly self-bound on surface. In order to reduce them, searching for very high mass NS is an essential method, since the maximum mass of NS determine the stiffness of EoS. Very recently, the discoveries of two $\sim 2M_{\odot}$ ruled out all soft EoSs \citep{Demorest10,Antoniadis13}, in which the predicted maximum masses of NS were lower than $2M_\odot$. On the other hand, searching for very low mass NS is also an attractive way. Because, the EoS of self-bound NS predicted distinct radii at low mass compared with ones by the EoS of gravity-bound NS. Moreover, gravity-bound NSs have minimum mass, while self-bound NSs do not. So, theoretical NS EoSs could be effectively tested from the accurate measurement of the radius for low mass NS.

\textit{EXOSAT} and \textit{RXTE} observed very low touchdown fluxes in PRE bursts from \object{4U 1746-37} \citep{Sztajno87,Galloway08}. During the cooling tail in its PRE bursts, the emission area remained near constant  \citep{Guver12a}. \citet{Sztajno87} assigned the peak fluxes as its Eddington flux. However, we assume that the Eddington luminosity was reached at the touchdown moment in \object{4U 1746-37's} PRE bursts similar as other sources. We also checked the persistent emission variations during X-ray bursts in \object{4U 1746-37}. The $f_a$-model do not provide better fitting results. After applying the Monto-Carlo simulation, we propose that a low mass NS ( $0.21\pm0.06~M_{\odot}$ or $0.41\pm0.14~M_{\odot}$, depends on accretion disc geometric effects) may exist in \object{4U 1746-37}.  Combined aboved two possibilities, the mass of \object{4U 1746-37} is $0.41^{+0.70}_{-0.30}~M_\odot$ at 99.7\% confidence. The peak fluxes in PRE bursts were not always consistent with touchdown fluxes. Two
geometric effects, the reflection of the far side accretion disc and the obscuration of the near side accretion disc were
possible attributed. In the case of accretion disc reflection, the derived mass and radius of NS in \object{4U 1746-37} could reproduced in the framework of self-bound NS EoSs, including quark-cluster stars and bare strange stars \citep{Lai09,Lai13,Guo13}.   In the
case of accretion disc obscuration, the self-bound NSs and gravity-bound NSs \citep{AP97,MPA87} are acceptable in 1-sigma and 2-sigma confidence level of the  mass and radius of NS in \object{4U 1746-37}, respectively. Three gravity-bound NS EoSs \citep{PAL88, MS96, Glendenning99} can be survival in 3-sigma confidence level.

\citet{Steiner10} discussed the possibility that the photosphere is still extended at the touchdown moment. In the extreme case, the Eddington flux is only dependent on the stellar mass. In Fig.~\ref{fig:dist3}, the contours of $M-R$ constrain the EoS same as Fig.~\ref{fig:dist1}. Again, self-bound NSs are acceptable in 1-sigma confidence level.   Two gravity-bound EoSs \citep{AP97,MPA87} and other three gravity-bound EoSs are possible in 2-sigma and 3-sigma confidence level.

Several low mass NSs (near or below $1M_\odot$) were also discovered in other binary systems, e.g., $1.07\pm0.36~M_{\odot}$ for \object{Her X-1} \citep{Rawls11}, $1.04\pm0.09~M_{\odot}$ for \object{SMC X-1} \citep{Meer07,Rawls11}, $0.87\pm0.07~M_\odot$ (eccentric orbit) or $1.00\pm0.01~M_{\odot}$ (circular orbit) for \object{4U 1538-52} \citep{Rawls11}, $0.72^{+0.51}_{-0.58}~M_{\odot}$ for \object{PSR J1518+4904} \citep{Janssen08}, however, without radii measurement.
A  low mass NS may be difficult to form from the collapse of a massive star. However, an extremely low mass of self-bound star (strange quark or quark-cluster star), even as low as planet-mass \citep{Xu03,Horvath12}, could exist through the accretion-induced collapse of a white dwarf \citep{Xu05,Du09}.


The equation of state of cold matter at supra-nuclear density, which is essentially related to the challenging non-perturbative behavior of quantum chromo-dynamics, is far beyond solved even nearly half a century after the discovery of pulsars. Based on different manifestations of pulsar-like compact stars (e.g., the featureless thermal X-ray spectrum and the free precession), it was conjectured by \citet{Xu2003} that pulsars could be so-called solid quark star, a kind of condensed objects composed of quark-clusters. The state of such quark-cluster matter is very stiff, and the resultant maximum mass of quark-cluster star would be even larger than $2M_\odot$ \citep{Lai09}, that is consistent with the later discoveries of massive pulsars \citep{Demorest10,Antoniadis13}. Additionally, pulsar glitches (sudden spin-up) can also be well understood in the regime of the quark-cluster star model \citep{Zhou14}.

In the conventional calculations of the crust of a strange star, one usually assumes that the bottom crust density could be as high as the drip density because the transmission probability through the Coulomb barrier is negligible for very heavy ions, e.g., A=118, Z=36 \citep{Alcock86}. However, accreted matter is mostly composed by ions not so heavy, and the transmission probability through the Coulomb barrier could be as high as $10^{-18}$ for $\rm ^{16}O$, according to the same approximations presented by \citet{Alcock86}. Normal matter accreted can then easily penetrate the Coulomb barrier, and thus can hardly exist outside a strange quark star (a new-born strange star could be bare because of strong exploding, otherwise a supernova might not be successful). Nevertheless, in the case of quark-cluster star, additional so-called strangeness barrier exists on quark-cluster surface. \citet{Xu14} demonstrated that a quark-cluster star may be surrounded by a hot corona or an atmosphere, and even a crust for different accretion rates, which could be
helpful to understand the O VIII Ly-$\alpha$ emission line in 4U 1700+24 \citep{Nucita14}. The mass of corona or atmosphere or crust is much less than the conventional value $\sim10^{-5}M_\odot$ of strange star, hence, the scale is negligible compared with the radius of NS.

On the other hand, \citet{Jaikumar06} suggested that the strange stars may have a neutralizing solid crust consisting of charged strangelets and electrons, if the surface tension is below the critical value of order a few $\rm{MeV/fm^2}$ \citep{Alford06}. \citet{Alford08} pointed out that the thickness of strangelet-crystal crust is sensitive to the EoS as well as the surface tension and can be changed from zero to hundreds of meters for a compact star of radius 10 km and mass 1.5$M_{\odot}$, or thicker for low mass NSs. But, the thickness of crust do not change the radius of NS significantly even in the extreme case of 4U 1746-37.

As demonstrated in this paper, the mass-radius curves of various quark-cluster stars and bare strange stars pass the case of \object{4U 1746-37}, no matter which geometric effects operated (reflection or obscuration). Certainly, our conclusions are based on the assumption that the observed PRE bursts were reached its Eddington luminosity. In future observation, if a brighter PRE burst is observed in \object{4U 1746-37}, then a larger mass NS is required. Moreover, we are expecting that the optical observations of next generation telescope TMT (Thirty Meter Telescope, http://www.tmt.org/) could provide rigorous mass constraints. With TMT, the optical light curves and spectroscopy could be capable to obtain the binary system information (such as inclination angle, the type of companion star, and mass function; \citealt{Antoniadis13}). And then, the mass of the compact object will be measured precisely and independently, which can verify the reliability of an ultra-low mass NS in \object{4U 1746-37}.

\section*{Acknowledgments}
We appreciate the referees' comments and suggestions, which improve our manuscript significantly. Z.S. Li thank Jean in 't Zand, Hauke Worpel and Duncan Galloway for discussion about \object{GRS 1747-312} observation. This work is supported by the 973 program No. 2012CB821801, the National Natural Science Foundation of China (11225314, 11173024), the Strategic Priority Research Program on Space Science of the Chinese Academy of Sciences (XDA04010300), XTP XDA04060604 and the Fundamental Research Funds of the Central Universities. This research has made use of data obtained from the High Energy Astrophysics Science Archive Research Center (HEASARC), provided by NASA's Goddard Space Flight Center.

\bibliographystyle{apj}

\newpage
\begin{deluxetable}{ccccc|c}
\tabletypesize{\scriptsize}
\tablecaption{PRE bursts in 4U 1746-37.\label{tb-1}}
\tablewidth{0pt}
\tablehead{
\colhead{Obs\_ID}    & \colhead{Touchdown flux } & \colhead{Peak flux} & \colhead{DCOR\tablenotemark{1}} &\colhead{PCU on\tablenotemark{2}}   & \colhead{$M$-$R$}\\
\colhead{}          & \colhead{$10^{-9}\rm erg/cm^2/s$}  &  \colhead{$10^{-9}\rm erg/cm^2/s$}     & \colhead{} &\colhead{}   &\colhead{}         }
\startdata
  30701-11-03-000 &   $2.86\pm0.16$ & $4.84\pm0.25$ &    1.028-1.035  & all&$0.21\pm0.06~M_{\odot}$, $6.26\pm0.99~{\rm km}$\tablenotemark{3} \\

  30701-11-04-00 &    $2.21\pm0.14$ & $5.23\pm0.26$ &    1.023-1.030  & all&$0.41\pm0.14~M_{\odot}$, $8.73\pm1.54~{\rm km}$\tablenotemark{4}  \\

  60044-02-01-03 &    $3.01\pm0.13$ & $5.84\pm0.23$ &    1.015-1.026  & 0,2,4& \\
\enddata
\tablenotetext{1}{Deadtime correction factor (DCOR) range. The exposure time of each burst spectrum is divided by DCOR.}
\tablenotetext{2}{The active Proportional Counter Units (PCUs) during the burst epoch.}
\tablenotetext{3}{The 1-sigma confidence level of mass and radius NS in 4U 1746-37, corresponding to Fig.~\ref{fig:dist1}.}
\tablenotetext{4}{The 1-sigma confidence level of mass and radius NS in 4U 1746-37, corresponding to Fig.~\ref{fig:dist2}.}
\end{deluxetable}

\end{document}